\documentclass[aps,amsmath,amssymb]{revtex4-2} 

\usepackage{graphicx}  
\usepackage{dcolumn}   
\usepackage{bm}        

\usepackage{verbatim}   
\usepackage{upgreek}

\usepackage[sort&compress]{natbib} 		

\usepackage[colorlinks=true,
			linkcolor=blue,
			citecolor=blue,
			urlcolor=blue]{hyperref}

\begin{document}

\title{Hydrodynamic Manipulation of Nano-Objects by Thermo-Osmotic Flows}
\author{Martin Fr\"anzl}
\author{Frank Cichos}
\email{cichos@physik.uni-leipzig.de}
\affiliation{Peter Debye Institute for Soft Matter Physics, Universit\"at Leipzig, Linnestr. 5, 04103 Leipzig, Germany}

\date{\today}
\begin{abstract}
The manipulation of micro- and nano-objects is of great technological significance to construct new materials, manipulate tiny amounts of liquids in fluidic systems, or detect minute concentrations of analytes. It is commonly approached by the generation of potential energy landscapes, for example, with optical fields. Here we show that strong hydrodynamic boundary flows enable the trapping and manipulation of nano-objects near surfaces. These thermo-osmotic flows are induced by modulating the van der Waals interaction at a solid-liquid interface with optically induced temperature fields. We use a thin gold film on a glass substrate to provide localized but reconfigurable point-like optical heating. Convergent boundary flows with velocities of tens of micrometres per second are observed and substantiated by a quantitative physical model. The hydrodynamic forces acting on suspended nanoparticles and attractive van der Waals or depletion induced forces enable precise positioning and guiding of the nanoparticles. Fast multiplexing of flow fields further provides the means for parallel manipulation of many nano-objects. Our findings have direct consequences for the field of plasmonic nano-tweezers as well as other thermo-plasmonic trapping schemes and pave the way for a general scheme of nanoscopic manipulation with boundary flows.
\end{abstract}

\maketitle

\section{Introduction} 

The control and manipulation of nano-objects is a key element for future nanophotonics \cite{Marago2013, Gao2017, Bradac2018, Xavier2018, Li2018}, material science \cite{Lin2017, Xavier2018, Xie2020}, biotechnology \cite{Gao2017, Favre-Bulle2019, Choudhary2019} or even quantum sensing \cite{Kayci2014}. Analytes dissolved in liquids, for example, need to be delivered, concentrated, separated or locally confined for further studies to become eventually processed and removed. Photonic elements including plasmonic nano-structures require precise positioning or controlled rearrangements to serve as adaptive functional structures. Key elements of the control at the micro- and nanoscale are often pressure driven fluidics transporting liquid volume and solutes as well as the generation of energy landscapes or force fields. The latter is achieved with optical \cite{Bradac2018} and plasmonic tweezers  \cite{Juan2011, Zhang2021}, magnetic fields \cite{DeVlaminck2012}, or using electrokinetic \cite{Wang2011} or opto-electronic \cite{Wu2011} effects. Especially in the field of plasmonic tweezers and nanoantennas where light is used to excite collective electron motion in noble-metals, the Joule losses lead to the unavoidable generation of heat at boundaries as an unwanted side effect  \cite{Kotnala2014, Jiang2020}. 
Yet, such optically generated temperature fields seem also suitable for the manipulation of nano-objects in liquids, for example, for the trapping of nanoparticles \cite{Braun2013} and single molecules \cite{Braun2015} or protein aggregates \cite{, Fraenzl2019} as well as for manufacturing active particles \cite{Jiang2009, Bregulla2014, Khadka2018, Fraenzl2021}. Those techniques rely on a drift of molecules and particles in optically generated temperature gradients termed thermophoresis or suggest thermo-electric effects  \cite{Lin2018} relying on a thermally induced charge separation. In addition, thermo-electrohydrodynamic effects using time-varying electric fields have been proposed for rapid particle transport \cite{Ndukaife2016, Hong2020} and convective effects that arise from temperature-induced density changes in the large liquid cells have been reported \cite{Donner2011, Roxworthy2014, Chen2020, Ciraulo2021}. 

Here we report on a fundamental physical process that is able to provide a versatile trapping and manipulation of nano-objects near surfaces in the simplest geometries. Contrary to most other techniques, our scheme is based on hydrodynamic flows generated by thermo-osmosis. Thermo-osmosis relies on a perturbation of the interfacial interactions at a solid-liquid boundary and is present in all experiments involving temperature gradients in plasmonic structures including plasmonic tweezers. We show that local temperature gradients on a thin gold film induce strong interfacial flows of several 10 to $100~\mathrm{\upmu m\,s^{-1}}$ in the direct vicinity (10 nm) of the gold film that results in a flow pattern reminiscent of convection. Based on a fully quantitative analysis of our experimental results we reveal that these thermo-osmotic flows on gold-water interfaces are induced by a temperature-induced perturbation of the van der Waals (vdW) interactions. Nano-objects suspended in the liquid are therefore dragged by the hydrodynamic forces originating from these flows. Utilizing attractive vdW interactions of the nano-object with the gold surface or temperature-induced depletion, we are able to trap and manipulate different types of nano-objects near the surface. Due to the speed of heating at small scales, we are able to multiplex flow fields to manipulate multiple objects with great precision. Our detailed analysis of the flow fields, the localization accuracy of nano-objects, and a comparison with numerical and theoretical predictions provide a quantitative understanding of these effects and paves the way for controlling boundary layer dynamics to manipulate objects at the smallest length scales in solutions. 

\begin{figure*}[!t]
\centering
\includegraphics[width=\linewidth]{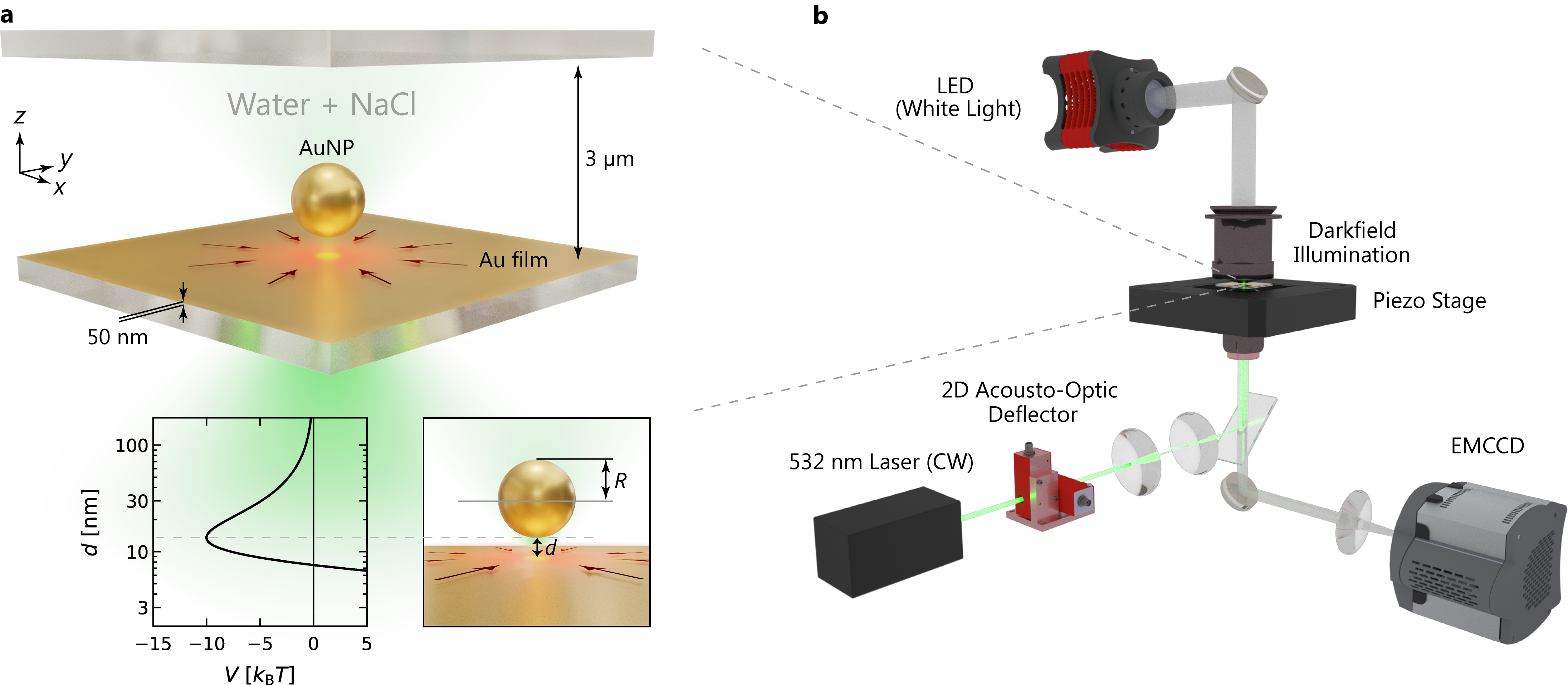}
\caption{\textbf{Thermo-hydrodynamic manipulation of Au NPs in NaCl solution.} \textbf{a}, Sketch of the sample design. It consists of two glass slides that confine a $5~\mathrm{\upmu m}$ thin liquid film of gold nanoparticles (AuNPs) dispersed in NaCl solution. The lower glass slide carries a 50~nm Au film that is locally heated by optical absorption a focused laser. \textbf{b}, Schematic of the experimental setup. It comprises an inverted optical microscope equipped with an acusto-optical deflector controlled, steerable focused laser with a wavelength of $\lambda = 532~\mathrm{nm}$. The AuNPs are observed using darkfield illumination with an oil-immersion dark-field condenser (NA 1.2) and a 100$\times$ oil-immersion objective set to NA 0.6 and recorded with an EMCCD camera.}
\label{Fig:Figure1}
\end{figure*}

\section{Results and Discussion} 

\subsection{Experimental Configuration and Working Principle} 

Our experiments rely on a simple sample geometry with a gold film (50 nm) that is deposited on a microscopy glass coverslip (Fig. \ref{Fig:Figure1}a). The sample chamber contains a suspension of gold nanoparticles (AuNPs) or other nano-objects (polystyrene NPs and ellipsoids) with a controlled amount of salt (NaCl),  surfactants (SDS, ...) or polymers (PEG). The gold film is heated locally in an inverted microscopy setup by a highly focused laser (532 nm) using beam steering optics (Acusto-Optic-Deflector, AOD). The nano-objects are observed using darkfield illumination with an oil-immersion darkfield condenser (NA 1.2) and a 100$\times$ oil-immersion objective set to NA 0.6 (Fig. \ref{Fig:Figure1}b). Additional details of the experimental setup and sample preparation are provided in the Methods section.

The trapping of nano-objects as detailed in the following is comprising two effects. i) The vertical confinement of the suspended objects as achieved by an attractive interaction of the suspended nano-objects with the gold surface, which is found to be the vdW interaction for gold nanoparticles and can be replaced by depletion forces for other materials.  ii) The generation of  thermo-osmotic boundary flows that are induced by the local heating and the corresponding perturbation of the liquid-solid interactions. This boundary flow is directed radially inwards to the heated spot and provides a confining hydrodynamic force on suspended objects at the heating spot.

\begin{figure*}[!t]
\centering
\includegraphics[width=\linewidth]{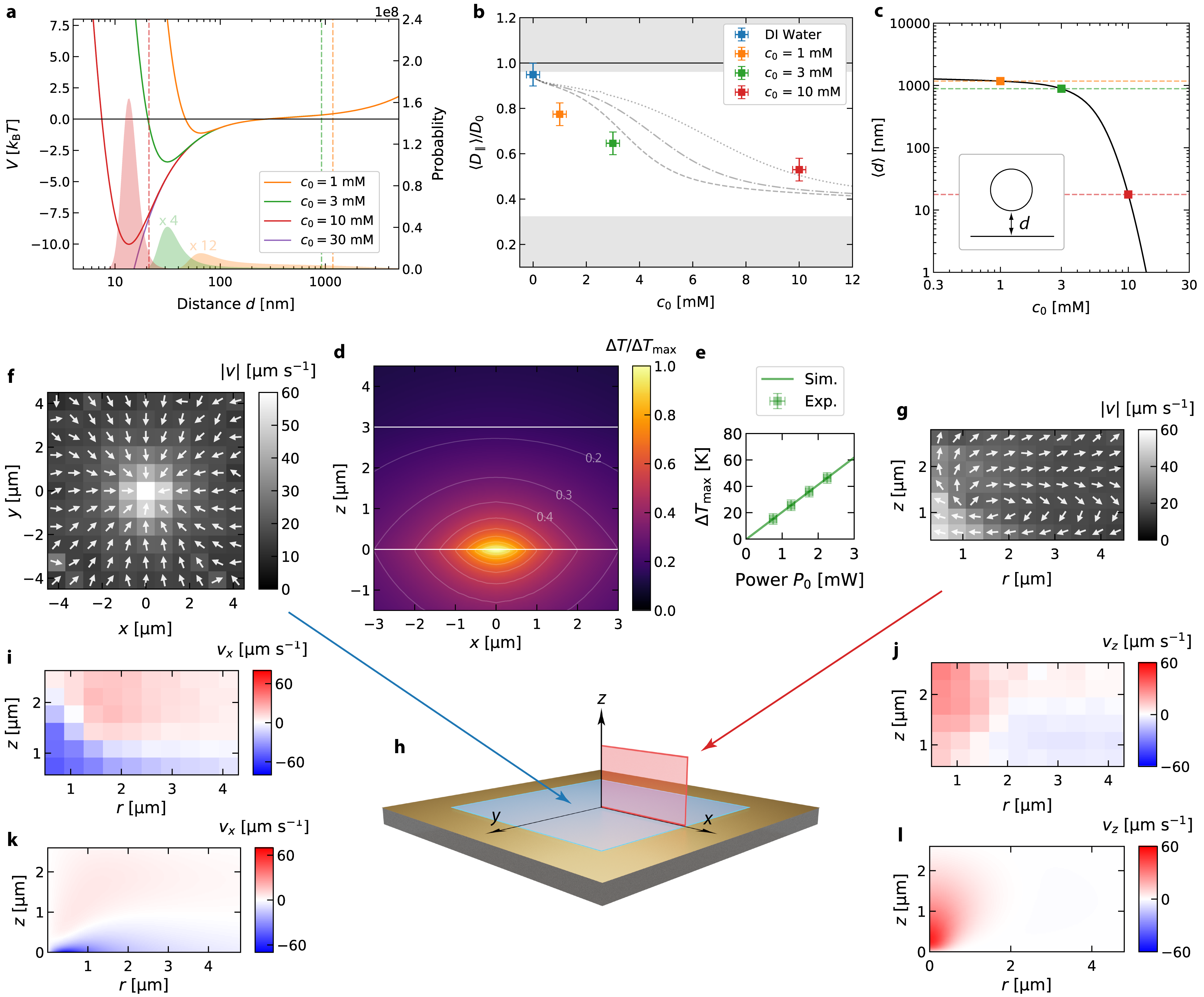}
\caption{\textbf{DLVO potential, lateral diffusion analysis, temperature distribution and thermo-osmotic flow field.} \textbf{a}, Plot of the DLVO potential, equation (\ref{Eq:TotalPotential}), between a 250~nm Au NP and a 50~nm Au film on a glass surface as function of particle-surface distance $d$ for different NaCl concentrations $c_0$. \textbf{b}, The measured diffusion coefficient $D_\parallel/D_0$ parallel to the Au film with respect to the bulk diffusion coefficient $D_0$ as function of the NaCl concentration. \textbf{c}, Relation between $\langle D_\parallel\rangle/D_0$ and the mean distance $\langle d\rangle$.  \textbf{d}, Simulation of the relative temperature increment in the $xz$-plane of the sample. \textbf{e}, Experimentally obtained temperature increment $\Delta T_\mathrm{max}$ as a function of the incident laser power $P_0$ (green data points) compared to the simulated values (green curve). \textbf{f}, Measured thermo-osmotic flow field in the $xy$-plane in close proximity to the gold film ($z < 500~\mathrm{nm})$. \textbf{g}, Measured thermo-osmotic flow field in the $xz$-plane. \textbf{h}, Illustration of the measured flow field planes in \textbf{f} and \textbf{g}. \textbf{i}, \textbf{j}, The $x$- and $z$-component of the measured flow velocities compared to the simulation results in \textbf{k} and \textbf{l}.}
\label{Fig:Figure2}
\end{figure*}

\subsection{Dynamics of AuNPs Close to a Au Film}

Consider a single AuNP with a radius of $R=125~\mathrm{nm}$ that is suspended in an aqueous solution of NaCl at a 10~mM concentration and diffusing in a thin liquid film of about $3~\mathrm{\upmu m}$ thickness over a 50~nm Au film (Fig. \ref{Fig:Figure1}a). Exploring the diffusion of the particle we observe a restriction of the $z$-positions to a thin layer close to the gold film. The gold particle never defocuses under these conditions while it does in deionized (DI) water (Supplementary Video~1). This restricted out-of-plane motion is the result of interactions comprising an attractive vdW contribution and a repulsion of the electrostatic double layers of the particle and surface \cite{Israelachvili2011} as described by the DLVO theory and the gravitational potential (see Supplementary Information for details). 
\begin{equation}
V(d, c_0) = V_\mathrm{E}(d, c_0) + V_\mathrm{vdW}(d) + V_\mathrm{G}(d)\ .
\label{Eq:TotalPotential}
\end{equation}
The total potential is depicted in Figure \ref{Fig:Figure2}a for different salt concentrations (see SI for parameters) as a function of the surface-to-surface distance $d = z - R$. The stronger screening of the surface charges at the gold film and the AuNP at higher salt concentration increase the importance of attractive vdW interactions to create this secondary minimum in the DLVO part of the potential. This potential also influences the observed dynamics as the particle couples with its hydrodynamic flow field to the solid boundary \cite{Lin2000}. The in-plane $D_{\parallel}$ (equation \ref{Eq:DiffusionCoefficientCloseToWall}) and out-of-plane $D_{\perp}$ diffusion coefficient (see Supplementary Information for details) are modulated with the distance $z$ of the particle from the wall.
\begin{equation}
\frac{D_\parallel(z)}{D_0} \approx 1 - \frac{9}{16}\frac{R}{z} + \frac{1}{8}\left(\frac{R}{ z}\right)^3 \pm\;\dots = \gamma_\parallel^{-1}(z) 
\label{Eq:DiffusionCoefficientCloseToWall}
\end{equation}
Over the course of a diffusion trajectory, the particle samples different diffusion coefficients according to its probability density $p(d) \propto \exp(-V(d,c_0)/(k_\mathrm{B} T))$ to be at a distance $d$ from the surface (filled regions in Fig. \ref{Fig:Figure2}a). The observed in-plane diffusion coefficient is thus a weighted average of the diffusion coefficient over the different vertical positions $d$. Using $p(d)$ we can calculate the corresponding salt concentration dependence of the diffusion coefficient and compare that to the experimental results. Figure \ref{Fig:Figure2}b shows that the experimentally observed $D_{\parallel}$ is decreasing with increasing salt concentration due to the hydrodynamic coupling in fair agreement with the theoretical predictions (see Supplementary Information for details). Calculating in addition the mean distance $\langle d \rangle$ of the particle from the surface reveals that only in the case of $c_0=10$ mM  the particle is confined in the DLVO potential well, while at lower salt concentrations an enhanced probability of finding the particle near the surface is the cause of the observed diffusion coefficient. These calculations help us to estimate the mean distance $\langle d \rangle$ of the particle from the surface, which is about $1.5~\mathrm{\upmu m}$ and $0.9~\mathrm{\upmu m}$ for the lowest NaCl concentrations (Fig. \ref{Fig:Figure2}c). At a concentration of $c_0 = 10$~mM the particle is hovering at a distance of $\langle d \rangle= 20$~nm surface. Note that this corresponds to values of $z/(2R) \approx 0.58$, which is far below the commonly explored region of the hydrodynamic coupling of colloids to walls \cite{Lin2000} allowing to experimentally explore new terrains also in the field of hydrodynamic wall coupling of colloids.

\subsection{Hydrodynamic Particle Confinement}

When tightly focusing the light of 532~nm wavelength to the gold film, a part of the incident energy (about 30~\%) is absorbed and converted into heat that perturbs the liquid-solid interactions. The temperature rise at the gold surface can be determined using a thin nematic liquid crystal (5CB) film and substantiated by finite element simulations with the complete three-dimensional temperature profile in the solution (see Fig. \ref{Fig:Figure2}d, e and Supplementary Information for details). 

These local temperature perturbations of the solid-liquid interactions at the interface induce a thermo-osmotic flow  \cite{Wurger2010, Bregulla2016}. Taking a liquid volume element close to the solid from the cold side and exchanging that with one at the hot side would not only transport heat since the liquid volumes have different temperatures, but also additional free energy as the liquid has a different interaction with the solid in these regions. The flow is induced in an ultrathin boundary layer corresponding in thickness to the length scale of liquid-solid interactions. Since the interaction range of liquid-solid interactions is only a few nanometers (the characteristic length of interfacial interactions), the boundary flow is due to its dimension collapsed into a quasi-slip hydrodynamic boundary condition:
\begin{equation}
v_\parallel = -\frac{1}{\eta}\int\limits_0^{\infty} z\,h(z)\;\mathrm{d}z\;\frac{\nabla_\parallel T}{T} = \chi\,\frac{\nabla_\parallel T}{T}\ ,
\label{Eq:SlipVelocity}
\end{equation}
where $h(z)$ is the excess enthalpy, $T$ the temperature and $\nabla_\parallel T$ is the temperature gradient parallel to the surface. The integral can be summarized to a thermo-osmotic coefficient $\chi$. The thermo-osmotic coefficient $\chi$, therefore, contains all information about the interfacial interaction between the liquid and the solid. If $\chi < 0$ the liquid is driven to the cold, whereas for $\chi > 0$, the liquid is driven to the hot. These boundary flows are present at all liquid-solid interfaces with tangential temperature gradients, though, they are commonly overlooked. They become particularly important for plasmonic and thermo-plasmonic trapping \cite{Kotnala2014}, as those techniques rely on the dynamics of molecules and particles in the direct vicinity of plasmonic nanostructures.
\begin{figure*}[t]
\centering
\includegraphics[width=\linewidth]{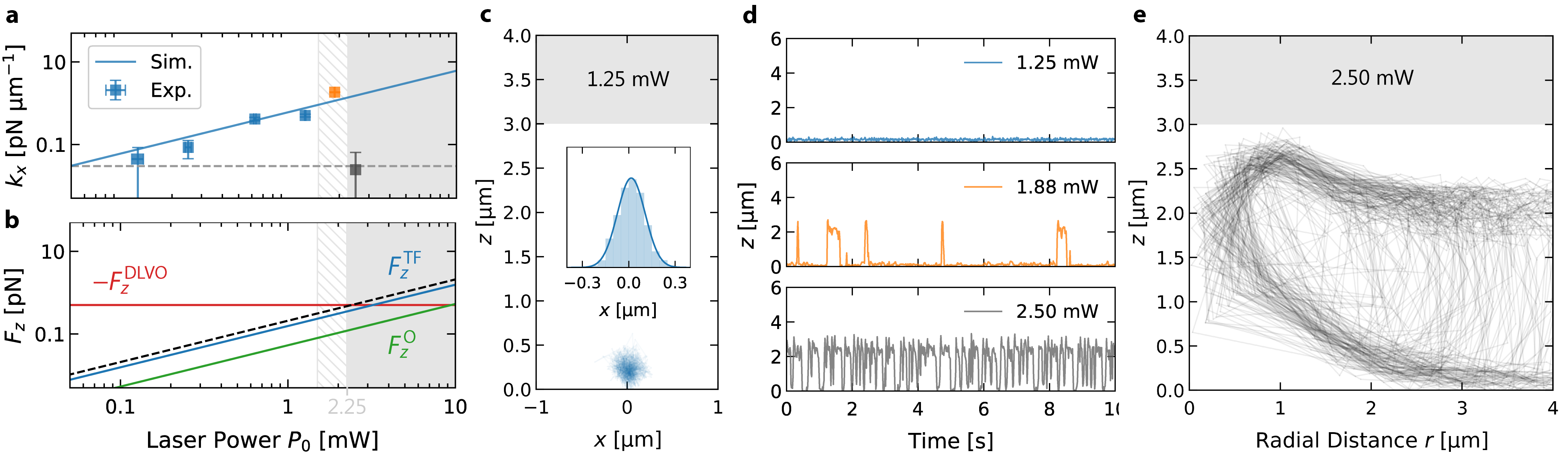}
\caption{\textbf{Forces on trapped NPs in 10~mM NaCl.} \textbf{a}, The lateral trap stiffness obtained from the experimental position histograms (blue data points) as function the laser power $P_0$ compared to the simulation result (blue solid line). \textbf{b}, The $z$-component of the thermo-osmotic drag force $F^\mathrm{TF}_z$ (blue line), the optical force $F^\mathrm{OF}_z$ (green line) and the total force $F^\mathrm{OF}_z + F^\mathrm{TF}_z$ (black dashed line) as function the incident laser power $P_0$ for a NP located at $x = 0$, $d = 30$ $(z = d + R)$. The attractive DLVO force $F^\mathrm{DLVO}_z$ is independent of the incident laser power and depicted as horizontal, red line. \textbf{c}, Trajectory of a AuNP for a heating laser power of 1.25~mW (see Supplementary Video~2 for details). The inset shows the corresponding lateral distribution histogram. \textbf{d}, Time traces of the $z$-position for three different laser powers. \textbf{e}, Trajectory of a AuNP for a heating laser power of 2.5~mW, this is, above the threshold power of 2.25~mW.}
\label{Fig:Figure3}
\end{figure*}
The boundary flow drives the flow field inside the fluid film. The resulting volumetric flow field can be tracked experimentally by single AuNPs in DI water, where the particles are not confined to a surface layer as reported above. We analyze the in-plane ($xy$) position of the particle and its $z$-position, where the latter is estimated from the radius $r_0$ of the defocussed particle images (see Supplementary Video~2 and Supplementary Information for details). The measured velocity distributions in the $xy$-plane near the gold layer and in the $xz$-plane are shown in Fig. \ref{Fig:Figure2}f and g, respectively. The $x$- and $z$-component of the measured flow velocities are depicted in Fig. \ref{Fig:Figure2}i, j and compare well to simulation results in Fig. \ref{Fig:Figure2}k, l. From these measurements, we extract a thermo-osmotic coefficient on the order of $\chi \sim 10\cdot 10^{-10}~\mathrm{m^2\,s^{-1}}$ (see Supplementary Information for details). We can break down the contributions to this value with equations (\ref{Eq:chiE}) and (\ref{Eq:chivdW}) to estimate the double layer and vdW contributions using the experimental parameters. Note that AuNP do not show thermophoresis due to their high thermal conductivity and thus isothermal surface. 
\begin{equation}
\chi_\mathrm{E} = \frac{\varepsilon\zeta^2}{8\eta} \approx 0.8\cdot 10^{-10}~\mathrm{m^2/s^{-1}}\ .
\label{Eq:chiE}
\end{equation}
For the electrostatic contribution we used $\zeta = -30~\mathrm{mV}$ \cite{Giesbers2002} and $\varepsilon = 80\,\varepsilon_0$ (see Methods section for details). An estimate of the vdW contribution can be given by
\begin{equation}
\chi_\mathrm{vdW} = \frac{A_\mathrm{H}\beta T}{3\pi\eta d_0} \approx 9.3\cdot 10^{-10}~\mathrm{m^2\,s^{-1}}\ ,
\label{Eq:chivdW}
\end{equation}
with $\beta = 0.2\cdot 10^{-3}~\mathrm{K}^{-1}$ being the thermal expansion coefficient of water and $d_0 = 0.2~\mathrm{nm}$ for the cut-off parameter \cite{Wurger2010} (see Supplementary Information for details). The sum of both contributions $\chi = \chi_\mathrm{E} + \chi_\mathrm{vdW} = 10.1\cdot 10^{-10}~\mathrm{m^2\,s^{-1}}$ matches well the experimental result and suggests that thermo-osmosis at gold-water interfaces is governed by vdW interactions. The obtained quasi slip velocities are ranging up to 80 $\upmu {\rm m/s}$ and provide, due to their omnipresence, a unique tool for nanofluidics. These thermo-osmotic flows are induced without any external pressure difference. They can be controlled by the light intensity heating laser and are quickly switched due to the extremely fast heat conduction at these length scales. Moreover the finding of the vdW dominated thermo-osmotic flows suggest that such contributions must be present in any plasmonic trapping experiment with extended gold structures \cite{Kotnala2014, Jiang2020, Hong2020, Zhang2021}. 

Using $F^\mathrm{TF}_x = 6\pi\eta R\,\gamma_\parallel\,v_x$ and $F^\mathrm{TF}_z = 6\pi\eta R\,\gamma_\perp v_z$, where $\gamma_\parallel$ and $\gamma_\perp$ are the correction factors for the friction coefficient of a sphere close to a surface we are able to extract the hydrodynamic forces that are exerted on the AuNP tracers (see equation (\ref{Eq:DiffusionCoefficientCloseToWall}) and Supplementary Information for details). The lateral forces allow to confine objects at the heating spot, yet the hydrodynamic force normal to the surface ($z$-direction) is repulsive without any additional interaction. Finally, such boundary flows with substantial vertical velocity gradients also exhibit a vorticity (see Supplementary Information for details) that generates a torque on suspended objects causing them to rotate  \cite{Bluemink2008}.

\begin{figure*}[b]
\centering
\includegraphics[width=\linewidth]{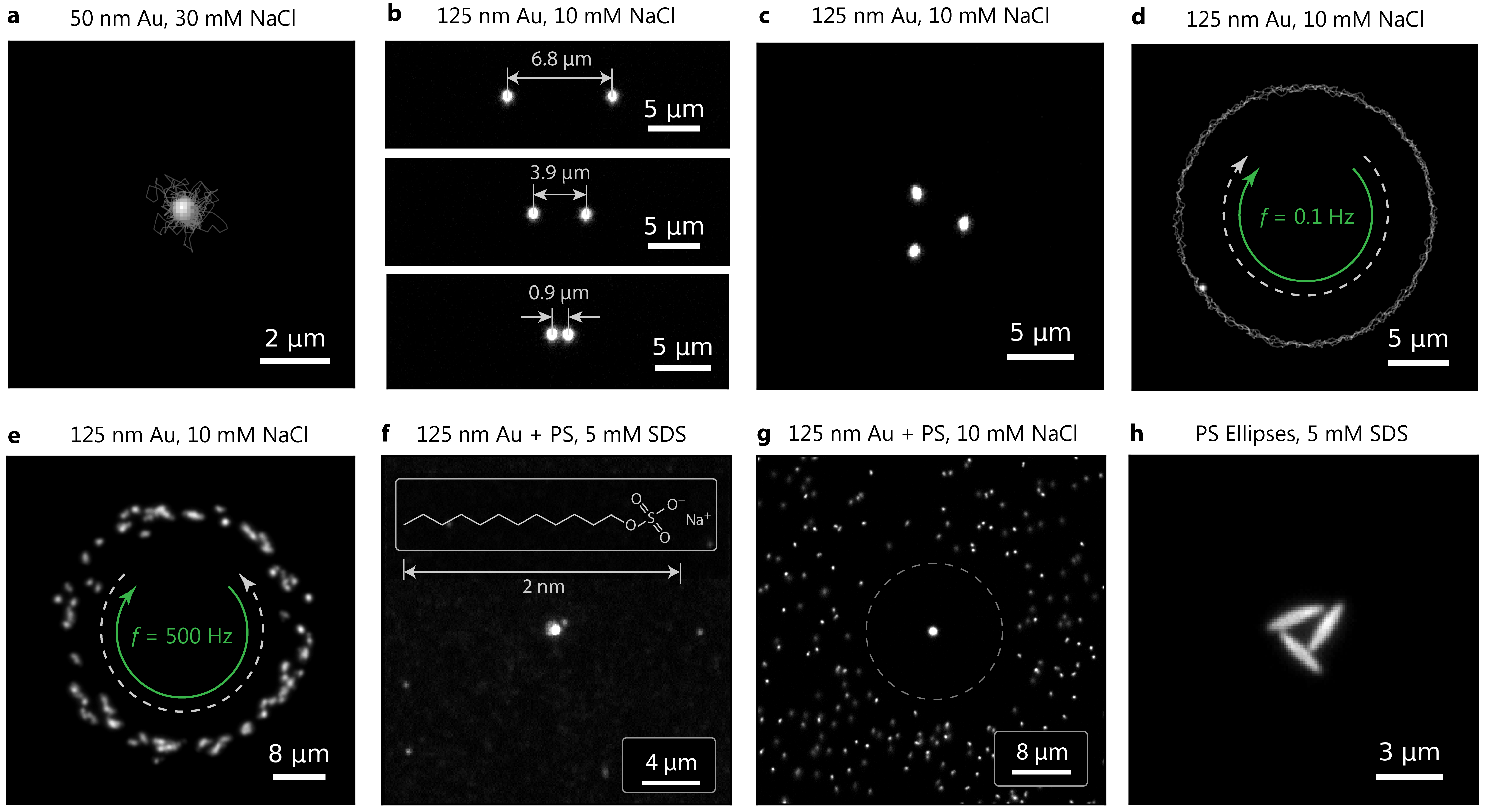}
\caption{\textbf{Manipulation of NPs over a Au film in NaCl and SDS solution.} \textbf{a}, A AuNP with 50~nm radius trapped at a NaCl of 30~mM (Supplementary Video~5). \textbf{b}, Manipulation of two AuNPs by a multiplexed laser beam (Supplementary Video~6). \textbf{c}, Control of three AuNPs (Supplementary Video~7). \textbf{d}, Actuation of a single AuNP on a circular trajectory by a steerable laser beam (Supplementary Video~8). The green and white, dashed arrows denote the moving direction of the laser focus and particle, respectively. \textbf{e}, Generation of thermo-viscous flows by rotating the laser focus on a circle with a rotation frequency of $f = 500~\mathrm{Hz}$ at high laser powers (Supplementary Video~9). Note, that laser movement (green arrow) and the thermo-viscous flow (white, dashed arrow) and have opposite directions. \textbf{f}, Attraction of a AuNP and PS NPs in 5~mM SDS due to depletion (Supplementary Video~10). \textbf{g}, A AuNP (125~nm radius) trapped in an ensemble of polystyrene (PS) NPs of the same size (Supplementary Video~11). \textbf{h}, Attraction of PS ellipsoids ($2.39~\mathrm{\upmu m}$ major-axis length, $0.34~\mathrm{\upmu m}$ minor-axis length) in 5~mM SDS (Supplementary Video~12). }
\label{Fig:Figure4}
\end{figure*}

\subsection{Single Particle Trapping and Flow Field Multiplexing}

The repulsive normal component caused by the hydrodynamic drag is now superimposed with an attractive force due to the DLVO potential when increasing the NaCl concentration. The surface-to-surface distance between AuNP and gold film and the depth of the appearing secondary DLVO potential minimum can be controlled by the NaCl concentration. At a NaCl concentration of about $c_0 = 10$~mM, the attractive potential has a depth of about $10\, k_\mathrm{B}T$ (see Fig. \ref{Fig:Figure2}a) and is strong enough to compete with the vertical drag force and additional optical forces on the AuNP to trap the particle above the heating spot.

Supplementary Video~3 demonstrates this trapping of an AuNP above the hot spot on the Au surface. This is purely the result of the hydrodynamic drag forces generated by the thermo-osmotic flow and the attractive vdW interaction between the AuNP and the Au film. This observation is substantiated by a quantitative evaluation of the lateral trap stiffness and vertical forces, as depicted in Fig. \ref{Fig:Figure3}a, b. The fluctuations of the particle in the hydrodynamic flow arise from a balance of the restoring hydrodynamic currents and the diffusive currents. Analysis of the lateral position histograms (inset in Fig. \ref{Fig:Figure3}c for 1.25~mW) yields an effective stiffness of the trap (Fig. \ref{Fig:Figure3}a) that well matches the predictions based on the thermo-osmotic flow (Fig. \ref{Fig:Figure2}i, j). The hydrodynamic trapping stiffness increases linearly up to a heating power of about $1.8$~mW. At this power, the vertical forces become strong enough to let the particle escape the secondary DLVO minimum, which is visible from the $z$-position time traces displayed in Fig. \ref{Fig:Figure3}d. The AuNPs are then observed to move vertically out of the DLVO potential to follow the flow inside the sample and to eventually return to the boundary flow via sedimentation (Fig. \ref{Fig:Figure3}e, Supplementary Video~4). The forces which eject the particle from the potential comprise the hydrodynamic and optical forces due to the radiation pressure from the heating laser leaked through the film. We have evaluated the individual contributions in simulations. They are shown together with the hydrodynamic force and the total vertical force as compared to the attractive force of the DLVO potential (Fig. \ref{Fig:Figure3}b) and provide quantitative agreement (threshold heating power of $2.25$~mW) with our experimental results. Note that while the stationary distribution of particles in the vertical direction is not influenced by the diffusive dynamics, the escape rate from the potential well is heavily altered by the fact that the vertical diffusion coefficient $D_{\perp}$ of the particle is decreasing to zero when approaching the gold film. This is enhancing the trapping times considerably (see Supplementary Information for details) but also increases the time required for the particle to enter the DLVO minimum by diffusion. 

The observed trapping is, hence, a vdW assisted thermo-hydrodynamic process. Vertical confinement is achieved by vdW attraction and double layer repulsion, while lateral confinement is the result of thermo-osmotic flows induced in an ultra-thin sheet of liquid at the interface. No additional contributions, for example, due to convective flows with similar flow patterns (see Supplementary Information for details) or thermo-electric effects are required for a quantitative description  \cite{Lin2016, Lin2017a, Lin2018, Ciraulo2021}. 

Precise tuning of the DLVO potential enables the trapping of even smaller Au NPs (Fig. \ref{Fig:Figure4}a, Supplementary Video~5). The speed of heat diffusion, which is about 4 orders of magnitude faster than the particle diffusion \cite{Baffou2020} allows us to introduce a flow field multiplexing. Therefore, we switch the heating location between different locations inducing flow thermo-osmotic flow fields for time periods of about 100 $\upmu {\rm s}$. With the help of this multiplexing, we are able to hold multiple $R = 125~\mathrm{nm}$ AuNPs (Fig. \ref{Fig:Figure4}b, c) at distances of less than 1 $\upmu {\rm m}$, which would not be possible with continuous heating of close-by locations (Supplementary Videos 6 and 7). A trapped AuNP can also be guided along the predefined path over the Au film as fast as $10~\mathrm{\upmu m\,s^{-1}}$ (Fig. \ref{Fig:Figure4}d, Supplementary Video~8). At larger manipulation speeds ($f > 100$~Hz) and higher heating power ($P_0 > 10$~mW) the thermo-osmotic attraction to the heating spot is combined with thermo-viscous flows \cite{Weinert2008, Weinert2011}. These flows originate from the temperature dependent viscosity $\eta(T)$ of the liquid and are directed opposite to the scanning direction of the laser \cite{Weinert2011}. The result of this combination of thermo-osmosis and thermo-viscous flows is a rotating ring-like particle structure (Fig. \ref{Fig:Figure4}e and Supplementary Video~9). These different effects that can be exploited in a simple planar geometry give rise to numerous applications including for example a freely configurable nanoparticle on mirror geometry for plasmonic sensing  \cite{Chikkaraddy2016, Li2018}.


\subsection{Beyond Thermo-Osmotic van der Waals Trapping}

So far, the presented manipulation is based on thermo-osmotic flows that drive the lateral motion of suspended colloids and a vertical confinement due to the secondary minimum of the DLVO potential between AuNP and Au film. While the thermo-osmotic flows are characteristic for all systems containing a heated gold/water interface including all previous studies on thermo-plasmonic trapping, the DLVO potential minimum is much weaker for other materials like polymer colloids or macromolecules due to their smaller vdW attraction. Often, those system even show a repulsion from the heat source due to thermophoresis, which is not present for AuNP. A more generalized strategy therefore needs additional attractive contributions, which confine suspended colloids or molecules to regions close to the gold surface to take advantage of the thermo-osmotic flow.  

Such attractive contributions can arise from depletion interactions \cite{Maeda2011, Jiang2009}. Thereby a temperature gradient repels dissolved molecules from the heated regions generating a concentration gradient that drives suspended nano-objects to the heating spot.
To demonstrate this effect we use the surfactant sodium dodecyl sulfate (SDS) at a concentration of 5 mM well below the critical micelle concentration (8.2~mM) to avoid complications of micelle formation. We suspend additional polystyrene particles(PS) and AuNPs of the same size ($R=125~\mathrm{nm}$) in the solution and compare their dynamics to a solution with AuNPs and PS particles without SDS but 10~mM NaCl. Remarkably, the heated spot is attractive for both AuNPs and for PS NPs (Fig. \ref{Fig:Figure4}f, Supplementary Video~10) in the SDS solution showing even PS colloidal crystal growth, while only the AuNP is trapped in the NaCl solution and the PS particles are repelled by thermophoresis (Fig. \ref{Fig:Figure4}g, Supplementary Video~11). 

The observations in NaCl are readily explained by the fact that the AuNP is confined in the DLVO minimum as demonstrated above but the PS particle is not due to a 10 times lower Hamaker constant (see Supplementary Information for details). The PS particle samples the whole liquid film thickness equally and not preferentially the region close to the Au film and experiences an additional thermophoretic drift velocity given by 

\begin{equation}
\vec u = - \frac{2}{3}\chi\frac{\nabla T}{T} = -D_\mathrm{T}\nabla T\ , 
\label{Eq:ThermophoreticVelocity}
\end{equation}

where $D_\mathrm{T}$ is the thermophoretic mobility \cite{Wurger2010} and $\nabla T$ the temperature gradient (Fig. \ref{Fig:Figure4}g, Supplementary Video~11). 
For $\chi > 0$ the particle is driven to the cold. From equation (\ref{Eq:chiE}) we find $\chi \approx \chi_\mathrm{E} = 1.28\cdot 10^{-10}~\mathrm{m^2\,s^{-1}}$ and $D_\mathrm{T} \approx 0.3~\mathrm{\upmu m^2\,K^{-1}\,s^{-1}}$, where we have used a measured zeta potential of $\zeta \approx -38~\mathrm{mV}$. The vdW contribution, $\chi_\mathrm{vdW}$ to either the thermophoretic drift or the attraction to the gold surface can be neglected due to the smaller Hamaker constant of PS. From the stationary probability distribution of the PS NP we find a Soret coefficient of $S_\mathrm{T} \approx 0.24~\mathrm{K}^{-1}$ (see Supplementary Information for details) in agreement with our theoretical prediction $S_\mathrm{T} = D_\mathrm{T}/D_\parallel \approx 0.21~\mathrm{K}^{-1}$.

In the SDS solution, the additional surfactant molecules now undergo thermophoresis to yield a concentration gradient in which suspended colloidal particles drift. The lower concentration in the heated regions promotes an effective attractive interaction of suspended colloids with the gold surface due to depletion forces. The drift velocity is described by an additional term to the thermodiffusion coefficient $D_\mathrm{T}$, that is, the second term in brackets in equation (\ref{Eq:DepletionVelocity})  \cite{Jiang2009, Wurger2010, Maeda2011}.
\begin{equation}
\vec u = -\left(D_\mathrm{T} - \frac{k_\mathrm{B}}{3\eta}R^2 c_0 N_\mathrm{A}\left(TS^\mathrm{SDS}_\mathrm{T} - 1\right)\right)\nabla T
\label{Eq:DepletionVelocity}
\end{equation}
Here $R$ is the size of the SDS molecule, $c$ the concentration in units of $\mathrm{mol/l}$ and $S^\mathrm{SDS}_\mathrm{T}$ the Soret coefficient of SDS. For $R = 2~\mathrm{nm}$ \cite{Syshchyk2016}, $c_0 = 5~\mathrm{mM}$ and $S^\mathrm{SDS}_\mathrm{T} = 0.03~\mathrm{K^{-1}}$~ \cite{Vigolo2010a} we find $-0.43~\mathrm{\upmu m^2\,K^{-1}\,s^{-1}}$ for the additional depletion contribution, which exceeds the thermophoretic mobility, $D_\mathrm{T} \approx 0.3~\mathrm{\upmu m^2\,K^{-1}\,s^{-1}}$, rendering the overall mobility negative. The PS NPs and the AuNPs are thus driven to the  the heated Au film surface(Fig. \ref{Fig:Figure4}f, Supplementary Video~10) which allows for further transport in the thermo-osmotic boundary flow. Additional contributions, as for example thermo-electric fields may even enhance the attractive components. 
Overall, this concept is readily transferred to other objects as shown in Figure \ref{Fig:Figure4}h and Supplementary Video~12, where we have trapped ellipsoidal PS particles in a 5~mM solution of SDS. 


\section{Conclusion} 

In conclusion, we have demonstrated that thermo-hydrodynamic boundary flows can manipulate nano-objects with unprecedented flexibility in a very simple sample geometry. These flows are the key for future thermo-optofluidic implementations with an extensive range of applications in the fields of  i) nanoparticle sorting and separation \cite{Xie2020}; ii) assembly of nanophotonic circuits \cite{Zhang2019} and plasmonic quantum sensors \cite{Li2018, Xavier2018}; iii) biotechnology on-chip laboratories \cite{Xavier2021} and iv) manufacturing of nanomaterials \cite{Lin2017, Xavier2018} and functional nanosurfaces \cite{Peng2019, Mohammadi2021}. We have substantiated our experimental findings of thermo-osmotic flow assisted trapping with a quantitative theoretical description. A flow field multiplexing scheme has been further developed to allow for the simultaneous manipulation of many individual nano-objects. Our concept can be combined with other thermally induced effects such as thermophoresis, depletion forces and thermoviscous flows to form a fully-featured nanofluidic system-on-a-chip. Besides direct consequences for the field of plasmonic nano-tweezers and other thermoplasmonic trapping schemes, the use of thermo-hydrodynamic flows as a tool for nanofluidic applications will extend the limits at the forefront of nanotechnology.

\bibliography{References.bib} 

\end{document}